\documentclass[aps,nofootinbib,showpacs,preprintnumbers,twocolumn,superscriptaddress,prb]{revtex4-1}

\usepackage{amsmath,amssymb}
\usepackage{graphicx}
\usepackage{bm}
\usepackage{color}
\usepackage{ulem}

\newcommand{\gbf}[1]{\boldsymbol #1}
\newcommand{\I}{\mathrm{i}}
\newcommand{\E}{\mathrm{e}}

\begin{document}

\title{$Z_2$ topological invariants in two dimensions from quantum Monte Carlo}

\author{Thomas C. Lang}
\affiliation{Department of Physics, Boston University, Boston, MA 02215, USA}
\affiliation{Institute for Theoretical Solid State Physics, RWTH Aachen University, Aachen, Germany}
\affiliation{JARA-HPC High Performance Computing}
\affiliation{JARA-FIT Fundamentals of Future Information Technology}
\author{Andrew M. Essin}
\affiliation{Department of Physics, University of Colorado, Boulder, CO 80309, USA}
\author{Victor Gurarie}
\affiliation{Department of Physics, University of Colorado, Boulder, CO 80309, USA}
\author{Stefan Wessel}
\affiliation{Institute for Theoretical Solid State Physics, RWTH Aachen University, Aachen, Germany}
\affiliation{JARA-HPC High Performance Computing}
\affiliation{JARA-FIT Fundamentals of Future Information Technology}

\begin{abstract}
We employ quantum Monte Carlo techniques to calculate the $Z_2$ topological invariant in a two-dimensional model of interacting electrons that exhibits a quantum spin Hall topological insulator phase. In particular, we consider the parity invariant for inversion-symmetric systems, which can  be obtained from the bulk's imaginary-time Green's function after an appropriate continuation to zero frequency. This topological invariant is used here in order to study the trivial-band to topological-insulator transitions in an interacting  system with spin-orbit coupling  and an  explicit bond dimerization. We discuss the accessibility and behavior of this topological invariant within quantum Monte Carlo simulations.
\end{abstract}

% http://www.aip.org/pacs/pacs2010/individuals/pacs2010_regular_edition/alpha_index.html
% 71.27.+a Strongly correlated electron systems
% 71.10.Fd Hubbard model electronic structure
% 71.30.+h Metal-insulator transition
% 73.43.-f Quantum Hall effects

\maketitle

%------------------------------------------------------------------------------------------------------------
\section{Introduction}
%------------------------------------------------------------------------------------------------------------
%
Topological insulators have been intensively explored in recent years~\cite{Hasan10,Qi11}, especially since their prediction~\cite{Bernevig06} and experimental realization in HgTe quantum wells~\cite{Koenig07}. Proposed also by Kane and Mele in a theoretical model of spin-orbit interactions in graphene~\cite{Kane05b} in search for an intrinsic quantum spin Hall (QSH) effect, topological band insulators in two dimensions are characterized in the presence of time-reversal invariance by a $Z_2$ topological index of the insulating electronic state~\cite{Kane05a}. In the case of non-interacting systems~\cite{Kane05a}, the $Z_2$ topological invariant can be extracted from the insulating band structure in analogy to the Thouless, Kohmoto, Nightingale and den Nijs (TKNN) classification of Block wave functions relevant for the integer quantum Hall effect~\cite{Thouless82}. Still in the context of non-interacting systems, it was found, that for inversion symmetric systems, e.g., in the sublattice-symmetric case on the graphene lattice, the $Z_2$ topological invariant can be easily extracted directly from the Hamiltonian matrix of the system at the so-called time-reversal invariant momenta (TRIM) in the Brillouin zone~\cite{Fu07,Hughes11,Turner12}. At these specific momenta, Kramers degenerate partners share the same band-structure eigenvalues, and from the parity of the occupied band eigenstates the corresponding $Z_2$ parity invariant (PI) is easily obtained. This approach will be reviewed below within a more general setting. 

Recently, topological insulators augmented with (strong) electron-electron interactions have attracted growing attention (see e.g. Ref.~\onlinecite{Hohenadler12b} for a recent review of  work on two-dimensional systems). Hence, the question arises, how the concept of a topological characterization of an insulating electronic state can be extended beyond the non-interacting band-structure regime. An important issue is, how such topological information can be efficiently calculated for interacting systems, in particular using unbiased numerical methods, such as quantum Monte Carlo (QMC) simulations. Several means to calculate topological invariants for interacting electronic insulators have been put forward~\cite{Volovik03,Wang10a,Gurarie11,Wang12a,Wang12b,Wang12c}. In a non-trivial generalization from the non-interacting case, these topological quantities are constructed based on the system's dressed single-particle Green's function, which remains a well-defined  quantity also for interacting systems. Of particular interest from a numerical perspective are the schemes presented in Refs.~\onlinecite{Wang12a,Wang12b}, which allow one to obtain the topological index based solely on the single-particle Green's function $G(\omega,\mathbf{k})$ at zero frequency $\omega=0$ and momentum $\mathbf{k}$. As will be shown  below, this quantity can be easily obtained from QMC calculations. The calculations can be further simplified for systems with explicit inversion symmetry, where $G(\omega=0,\mathbf{k})$ needs to be obtained at the TRIM only~\cite{Wang12b}, similar to the non-interacting case~\cite{Fu07}. A description of this approach to extract the PI for interacting systems, which in addition also exhibit spin $S^z$ conservation, will be presented below.

Such methods to obtain topological invariants for interacting systems have been applied recently, e.g., to correlated electron systems in one dimension using the numerically exact time-dependent density matrix renormalization group (DMRG) approach~\cite{Manmanna12}. For two-dimensional interacting systems, approximate means to estimate the Green's function have been employed; for example the variational cluster approximation (VCA) has been applied to the Kane-Mele model with local interactions (the Kane-Mele-Hubbard model~\cite{Rachel10,Zheng11,Hohenadler11,Wu12,Hohenadler12a,Assaad13})~\cite{Budich12a} to study the transition from the quantum spin Hall topological insulating phase to the antiferromagnetic Mott insulator regime at strong interactions. Dynamical mean-field theory (DMFT) has been employed to study the interaction-driven transition between topological states in a Kondo insulator \cite{Werner13} and cluster DMFT to study the three dimensional pyrochlore iridates \cite{Go12}.

Here, we set out to employ unbiased and numerically exact methods to access Green's function-based topological invariants in two dimensional fermion systems. In particular, we  use a projective QMC scheme to study the PI for a Kane-Mele-Hubbard model with anisotropic hopping, which exhibits a topological insulator regime, a trivial, non-magnetic insulating phase, as well as an antiferromagnetically ordered Mott insulating regime. We analyze the PI in these phases, the transitions between them and assess the PI's characterization of these different regimes. The goal of this paper is not to provide a detailed analysis of the complete phase diagram of this model, but to instead illustrate the actual application of the Green's function approach to study topological invariants in interacting two-dimensional fermion systems. 

The rest of this paper is organized as follows: In the next section, we introduce the dimerized Kane-Mele-Hubbard model that we explore further below, followed by a review of how to extract the PI from the zero-frequency Green's function. We examine the non-interacting limit of our model, where the Green's function and the PI may be easily calculated. After that, we discuss how to obtain the PI for finite interactions from QMC simulations, before we then apply this approach to the dimerized Kane-Mele-Hubbard model.

%------------------------------------------------------------------------------------------------------------
\section{Dimerized Kane-Mele-Hubbard model}
%------------------------------------------------------------------------------------------------------------
%
\begin{figure}[t!]
\centering
  \includegraphics[width=\columnwidth]{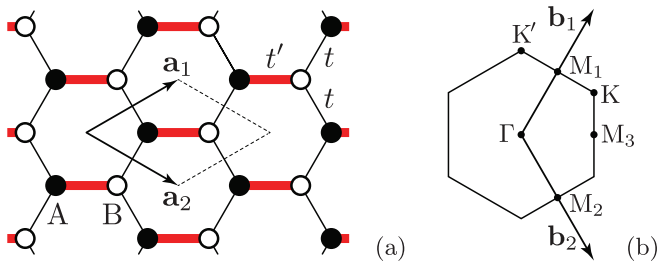}
  \caption{(Color online) (a) Honeycomb lattice with a unit cell indicated by dashed lines. The arrows indicate the lattice vectors $\mathbf{a}_{1,2}$. Bold (red) lines denote bonds with hopping amplitude $t'$, while the hopping amplitude along the other bonds on the honeycomb lattice equals $t$. Filled (open) circles indicate lattice sites belonging to the A (B) sublattice. (b) Brillouin zone with the time-reversal invariant momenta $\{\Gamma, \text{M}_1, \text{M}_2, \text{M}_3\}$, and the reciprocal lattice vectors $\mathbf{b}_{1,2}$ indicated. 
  \label{fig:lattice}}
\end{figure}

In the following, we consider the half-filled Kane-Mele-Hubbard model~\cite{Rachel10,Zheng11,Hohenadler11,Hohenadler12a} with an additional, explicit bond dimerization, described by the Hamiltonian
\begin{equation}
 H=H_0+H_\text{SO}+H_U,
\end{equation}
with the nearest-neighbor hopping terms
\begin{equation}
H_0=-t'\sum_i \sum_\sigma(a^\dagger_{i\sigma} b_{i\sigma} + h.c.)
    -t\sum_{\langle i,j \rangle}\sum_\sigma (a^\dagger_{i\sigma} b_{j\sigma} + \mathrm{h.c.}) \,,\nonumber
\end{equation}
the spin-orbit next-nearest-neighbor term
\begin{equation}
H_\text{SO}= \I\, \lambda\sum_{\langle\langle i,j \rangle\rangle} \nu_{ij} (a^\dagger_{i\sigma} \sigma^z_{\sigma\sigma'} a_ {j\sigma'} +  b^\dagger_{i\sigma} \sigma^z_{\sigma\sigma'} b_{j\sigma'}) \,,\nonumber
\end{equation}
and the Hubbard local interaction term
\begin{equation}
H_U=U\sum_i (a^{\dagger}_{i\uparrow}a_{i\uparrow}a^{\dagger}_{i\downarrow}a_{i\downarrow} + b^{\dagger}_{i\uparrow}b_{i\uparrow}b^{\dagger}_{i\downarrow}b_{i\downarrow}) \,.\nonumber
\end{equation}
Here, $a^\dagger_{i\sigma}$ ($b^\dagger_{i\sigma}$) denote creation operators for  spin-$\sigma$ fermions on a sublattice-A site (sublattice-B site), with $i$ denoting the two-site unit cell at position $\mathbf{r}_i$ on the honeycomb lattice. The spin-orbit coupling strength is denoted by $\lambda$, while $\nu_{ij}=\pm 1$ depending on whether the considered hopping process involves a left, or a right turn. We allow for a different nearest-neighbor hopping strength $t'$ along one of the three nearest-neighbor bond directions, as compared to the other directions, cf. Fig.~1. Here, the unit cell is chosen such that it contains a $t'$ bond and is centered on this bond. The two lattice vectors of the honeycomb lattice $\mathbf{a}_{1,2}= a_0 (3/2,\pm \sqrt{3}/2)$ are also shown in Fig.~1(a). In the following, we set the distance between nearest neighboring lattice sites $a_0=1$. For $t'=t$, the usual Kane-Mele-Hubbard model is recovered, which for finite spin-orbit coupling $\lambda$ and in the small-$U$ regime features a QSH topological insulating region, adiabatically connected to the $U=0$ QSH state. Increasing the onsite repulsion $U$ eventually drives the system into an ordered phase with long-ranged transverse antiferromagnetic correlations~\cite{Hohenadler11}. Furthermore, at $U=0$, the explicit bond dimerization allows to drive the system from the topological insulator QSH state to a (trivial) band insulating phase for $t'>2t$. At $t'/t=2$, the system is gapless with the bulk gap closing at one of the M-points in the Brillouin zone, cf. Fig.~1(b). This will be examined in more detail below as well as the properties of the model for $t'>t$ and finite interactions, $U>0$. To study the effects of interactions in terms of the topological invariants, we employ quantum Monte Carlo simulations to calculate the imaginary-time Green's functions of this model Hamiltonian and then transform to the Green's function at zero frequency, from which we extract the $Z_2$ PI for this inversion symmetric system.

%------------------------------------------------------------------------------------------------------------
\section{Parity invariant from Green's function}\label{Sec:PI-from-GF}
%------------------------------------------------------------------------------------------------------------
%
In an inversion symmetric system, the PI may be calculated from the system's Green's function following Ref.~\onlinecite{Wang12a}, which generalizes the procedure from the non-interacting case~\cite{Fu07}. Here, due to the explicit $S^z$ conservation of the Hamiltonian, the Green's function is block-diagonal in spin-space, and the procedure can be restricted to a single spin sector. The zero-frequency Green's function for each spin sector, $G_{\sigma}(0,\mathbf{k})$, where $\sigma=+1$ ($-1$) for  spin-up (spin-down), thus is a $2\times 2$ matrix in the A/B-sublattice basis. Denoting by $\mathbf{b}_{1,2}$ the reciprocal lattice vectors (with $\mathbf{b}_i\cdot \mathbf{a}_j=2\pi \delta_{ij}$), we consider the four TRIM
\begin{equation}
   \gbf{\kappa}_{n_1,n_2}=n_1\mathbf{b}_1/2+n_2\mathbf{b}_2/2, \quad n_i=0,1 \,,
\end{equation}
corresponding to the $\Gamma$-point and the three M-points indicated in Fig.~1(b), at which the operation of inversion commutes with the zero-frequency Green's function $G_{\sigma}(0,\mathbf{k})$. Here, the operation of inversion that interchanges the two sublattices and squares to the identity can be represented in the sublattice basis by the first Pauli-matrix,  $P=\sigma^x$. Simultaneously diagonalizing the two matrices $P$ and $G_\sigma(0,\gbf{\kappa}_{n_1,n_2})$, we identify for each of the four TRIM the eigenvalue of $P$ for the common eigenvector with a positive eigenvalue of $G_\sigma(0,\gbf{\kappa}_{n_1,n_2})$. These eigenvectors are referred to as right-zeros or R-zeros in Ref.~\onlinecite{Wang12a}.  Denoting the corresponding $P$ eigenvalue of the R-zero by $\eta_{\gbf{\kappa}_{n_1,n_2}}$, we obtain the PI, $\Delta\in\{0,1\}$, as
\begin{equation}
   (-1)^\Delta=\prod_{n_1,n_2} \eta_{\gbf{\kappa}_{n_1,n_2}} \,,
\end{equation}
from any of the two spin sectors, which together form a Kramer's pair at each TRIM. It is thus sufficient for the calculation of the PI, to only consider e.g. the spin-up sector due to the explicit $S^z$ conservation of the Hamiltonian. The procedure is however easily generalized to inversion symmetric systems without explicit $S^z$ conservation~\cite{Wang12a}.

%------------------------------------------------------------------------------------------------------------
\section {Non-interacting case}\label{Sec:non-interacting}
%------------------------------------------------------------------------------------------------------------
%
To illustrate the above procedure, let us first consider the non-interacting limit, i.e., the dimerized Kane-Mele model.  
For $U=0$, the Hamiltonian $H$ can be diagonalized directly via a transformation to momentum space, 
\begin{equation}
   H=\sum_{\mathbf{k},\sigma} (a^\dagger_{\mathbf{k},\sigma} \: b^\dagger_{\mathbf{k},\sigma})\; h_{\sigma}(\mathbf{k}) \left(\begin{matrix}a_{\mathbf{k},\sigma} \\ b_{\mathbf{k},\sigma}\end{matrix}\right).
\end{equation}
In each spin sector $\sigma=+1$ ($-1$), the Hamiltonian matrix at wave vector $\mathbf{k}$ equals
\begin{equation}
   h_{\sigma}(\mathbf{k})=\left(\begin{matrix}
                     \sigma \gamma_\mathbf{k} & -g_\mathbf{k} \\
                      -g^*_\mathbf{k} & -\sigma\gamma_\mathbf{k}
                    \end{matrix}\right),
\end{equation}
where $g_\mathbf{k}=t'+t (\E^{\I\mathbf{a}_1\cdot\mathbf{k}}+ \E^{\I\mathbf{a}_2\cdot\mathbf{k}})$ relates to the nearest neighbor hopping terms and $\gamma_\mathbf{k}=2\lambda(-\sin(\sqrt{3} k_y)+2 \cos(3k_x/2) \sin(\sqrt{3} k_y/2))$ to the spin-orbit term.  The system described by $H$ conserves $S_z$, such that the Green's function $G(\omega,\mathbf{k})$ is block-diagonal in spin-space, and each spin component in the non-interacting case equals
\begin{equation}\label{eq:greenfree}
    G_{\sigma}(\omega,\mathbf{k})=\left[\omega-h_{\sigma}(\mathbf{k})\right]^{-1}.
\end{equation}
At zero frequency this is essentially the inverse of the Hamiltonian matrix:
\begin{equation}\label{eq:gf0U0}
   G_{\sigma}(0,\mathbf{k})=-h^{-1}_{\sigma}(\mathbf{k}) \,.
\end{equation}
Based on the approach outlined in the previous section, we then obtain for finite values of $\lambda$ a change in the PI from ${\Delta=1}$ for ${t'<2t}$ to ${\Delta=0}$ for ${t'>2t}$. This indicates the change from a topological insulator to a trivial band insulating state driven by the explicit bond dimerization. At ${t'=2t}$, the system becomes semi-metallic due to the single particle gap closing at the M$_3$ point, i.e., at ${\mathbf{k}=\gbf{\kappa}_{1,1}}$. This can be seen from the band structure shown for ${\lambda/t=0.2}$ in Fig.~2.
\begin{figure}[t!]
\centering
  \includegraphics[width=\columnwidth]{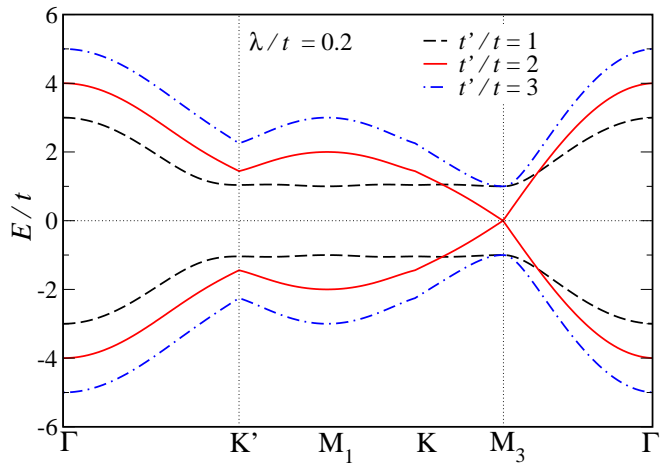}
  \caption{(Color online) Band structure of the dimerized Kane-Mele model along the indicated path through the Brillouin zone for $\lambda/t=0.2$ and different values of $t'$, as indicated. 
  \label{fig:bs}}
\end{figure}
In the following, we will examine this transition also at finite values of $U$. Before performing such an analysis, we first explain, how we extract the PI in the interacting regime from QMC  simulations.

%------------------------------------------------------------------------------------------------------------
\section{Parity invariant from QMC}
%------------------------------------------------------------------------------------------------------------
%
Once the zero-frequency Green's function $G(0,\mathbf{k})$ has been obtained for the interacting model, the PI can be calculated as outlined in Sec.~\ref{Sec:PI-from-GF}. In analogy with Eq.~(\ref{eq:gf0U0}) for the non-interacting case, one can  associate to the interacting model a fictitious Hamiltonian matrix $h_\text{topol}(\mathbf{k})=-G^{-1}(0,\mathbf{k})$, which has been dubbed the topological Hamiltonian~\cite{Wang12c}. It contains the topological information of the interacting model, where for the free case $h_\text{topol}(\mathbf{k})$ equals the Hamiltonian matrix of $H$. Hence, we merely need to consider, how the zero-frequency Green's function is obtained from the QMC calculations. In particular, we employed a projective QMC scheme, by which we obtain the momentum and spin resolved single particle Green's function in imaginary time within the system's ground state on finite lattices. To obtain the PI, we then  calculate from the imaginary-time data of the Green's function those at Matsubara frequencies, and continue in particular to zero frequency. For this purpose, let us first consider the system at a finite temperature $T=1/\beta$. The imaginary time Green's function $G_\sigma(\tau,\mathbf{k};\beta)$ at a given momentum $\mathbf{k}$ and spin projection $\sigma$ is a two-by-two matrix with entries
\begin{equation}
   [G_\sigma(\tau,\mathbf{k};\beta)]_{jl}=-\langle c_{\mathbf{k},\sigma,j}(\tau)\, c^\dagger_{\mathbf{k},\sigma,l}(0) \rangle_\beta \,,
\end{equation}
where $j,l=1,2$ is a sublattice index, with $c_{\mathbf{k},\sigma,1}=a_{\mathbf{k},\sigma}$ and $c_{\mathbf{k},\sigma,2}=b_{\mathbf{k},\sigma}$. For frequencies $\omega_n=2(n+1)\pi/\beta$ the Matsubara-Green's function is then given as
\begin{equation}
   G_{\sigma}(\I\omega_n,\mathbf{k};\beta)=\int_0^\beta G_{\sigma}(\tau,\mathbf{k};\beta)\, \E^{\I \omega_n \tau} d\tau \,. 
\end{equation}
Particle hole symmetry of the model at half-filling, i.e., under the transformation ${c^\dagger_{\mathbf{k},\sigma,j}\rightarrow d_{\mathbf{k},\sigma,j} = (-1)^j c^\dagger_{-\mathbf{k},\sigma,j}}$ in each spin sector together with inversion symmetry leads to the following conditions on $G_{\sigma}(\tau,\mathbf{k};\beta)$: For equal sublattices, $[G_\sigma(\tau,\mathbf{k};\beta)]_{jj}=[G_\sigma(\beta-\tau,-\mathbf{k};\beta)]_{jj}$, while, for $j\neq l$, ${[G_\sigma(\tau,\mathbf{k};\beta)]_{jl}=-[G_\sigma(\beta-\tau,-\mathbf{k};\beta)]_{jl}}$.

We thus obtain for the diagonal elements of the Green's function at one of the TRIM $\gbf{\kappa}=\gbf{\kappa}_{n_1,n_2}$ the equation
\begin{equation}
   [G_{\sigma}(\I \omega_n,\gbf{\kappa};\beta)]_{jj}=2\,\I\int_0^{\beta/2} [G_\sigma(\tau,\gbf{\kappa};\beta)]_{jj} \sin(\omega_n \tau)\, d\tau \,,
\end{equation}
and, for $j\neq l$, 
\begin{equation}
   [G_{\sigma}(\I \omega_n,\gbf{\kappa};\beta)]_{jl}=
      2 \int_0^{\beta/2} [G_\sigma(\tau,\gbf{\kappa};\beta)]_{jl} \cos(\omega_n \tau)\, d\tau \,.
\end{equation}
Now, the limit $\beta\rightarrow\infty$ can be taken properly:  From the projective QMC, we obtain the ground state Green's function $G_\sigma(\tau,\gbf{\kappa})=\lim_{\beta\rightarrow\infty}G_\sigma(\tau,\gbf{\kappa};\beta)$, and then perform the above integrals with a sufficiently large cutoff $\beta\rightarrow \theta$, set e.g. by the imaginary time evolution length of the Green's function $\theta$ employed in the QMC simulations.
Here, we used $\theta=20/t$. This cutoff proved to be sufficient for the GreenÕs function to decay to zero within error bars, especially for large values of $U/t$, but for the extreme cases close to the topological-to-trivial band insulator transition, where the gap becomes very small. Note, that one cannot simply take the limit $\I\omega_n\rightarrow 0$ before accounting for the (anti)symmetry conditions on the imaginary time Green's functions. This would lead to wrong results, as exemplified below. After (anti)symmetrization, the limit $\I\omega_n\rightarrow 0$ can be performed with the $T=0$ Green's functions, so that in particular,
\begin{equation}
   [G_{\sigma}(\omega=0,\gbf{\kappa})]_{jj}=0 \,,
\end{equation}
and, for $j\neq l$, 
\begin{equation}\label{eq:gomegajl}
   [G_{\sigma}(\omega=0,\gbf{\kappa})]_{jl}= 2\int_0^{\theta/2} [G_\sigma(\tau,\gbf{\kappa})]_{jl}\, d\tau \,.
\end{equation} 
Hence, within the QMC simulations, one merely needs to measure the off-diagonal part of the Green's function explicitly. To illustrate the above point, consider for a moment the non-interacting limit, for which the exact $T=0$ imaginary-time Green's function
\begin{equation}
   G_{\sigma}(\tau,\gbf{\kappa})= -\frac{1}{2}\, \E^{-|g_{\gbf{\kappa}}|\tau}\left(
   \begin{matrix}
      1 & -1\\
     -1 &  1
   \end{matrix}
\right).
\end{equation}
If calculated naively, via ${\int_0^\infty G_{\sigma}(\tau,\Gamma) \,d\tau}$, one would (wrongly) obtain a finite value of $[G_{\sigma}(\omega=0,\Gamma)]_{jj}$ instead of the actual value (i.e. zero), which also follows in this case directly from Eq.~(\ref{eq:greenfree}).

%------------------------------------------------------------------------------------------------------------
\section{QMC results}
%------------------------------------------------------------------------------------------------------------
%
After having examined the calculation of the PI for the interacting system in the previous section, we now present results from QMC simulations of the dimerized Kane-Mele-Hubbard model. We employ a projector axillary-field determinantal QMC scheme\cite{AsEv08} by which we obtain the momentum and spin resolved single particle Green's function in imaginary time within the system's ground state for finite lattices with $N=2L^2$ lattice sites employing periodic boundary conditions. Here, $L$ denotes the linear system size, which for multiples of six allows all TRIM as well as the corners of the Brillouin zone (the so-called Dirac points) to be presented. In particular, we use a projection length $\Theta=50/t$, imaginary-time step $\Delta\tau=0.05/t$ and linear systems sizes $L=6$, $12$ and $18$. An imaginary time evolution length $\theta=20/t$ has been used to obtain the Green's function, as discussed in Sec.~V. Details on the employed QMC method in application to the Kane-Mele-Hubbard model can be found in Ref.~\onlinecite{Hohenadler12a}.

\begin{figure}[t!]
  \centering
  \includegraphics[width=\columnwidth]{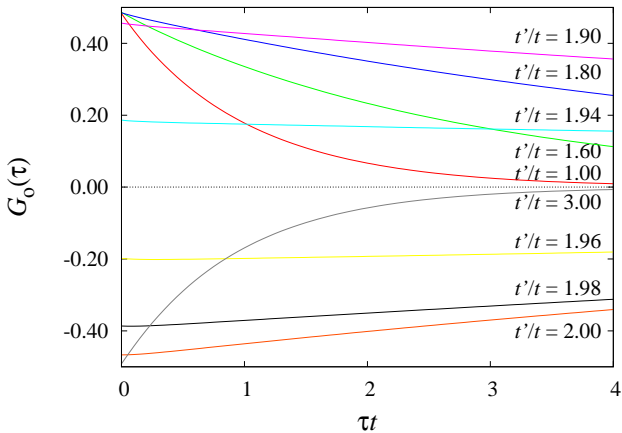}
  \caption{(Color online) Off-diagonal component of the Green's function at the M$_3$-point for $L=6$, $U/t=2$ and $\lambda/t=0.2$ at various values of $t'/t$. Error bars are of the order of the line width and have been omitted for clarity. 
  \label{fig:GtauM3_U2}}
\end{figure}

\begin{figure}[t!]
\centering
  \includegraphics[width=\columnwidth]{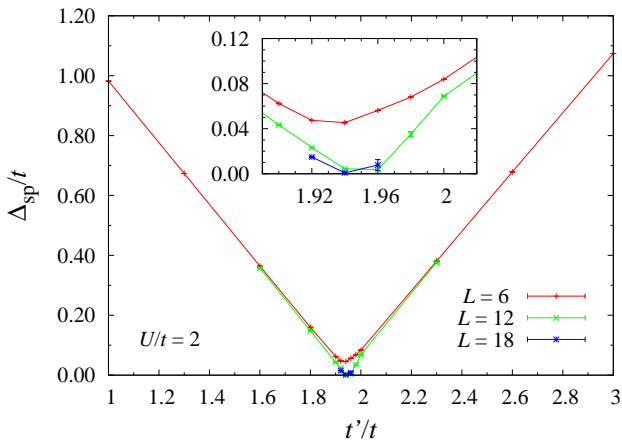}
  \caption{(Color online) Evolution of the single particle gap $\Delta_{\mathrm{sp}}$ for different system sizes as a function of $t'/t$ for $U/t=2$ and $\lambda/t=0.2$
near the quantum phase transition from the topological insulator to the dimerized phase. The inset focuses in on the transition region. 
  \label{fig:deltasp}}
\end{figure}

To test the feasibility of extracting the PI within QMC, we first consider the $t'/t$-driven transition between the topological insulator regime and the trivial band insulator for large $t'$ at finite values of $U$. In the following, we consider $\lambda/t=0.2$, in order to focus on the QSH to dimerized insulator transition without being compromised by the influence of the QSH-insulator transition at $\lambda=0$, and without loss of generality.\cite{Meng10,Sorella12,Hohenadler11,Hohenadler12a} As an example, Fig.~\ref{fig:GtauM3_U2} shows the imaginary-time dependence of the off-diagonal component of the Green's function at the M$_3$-point at $\gbf{\kappa}_{11}$, which in the following we denote by
\begin{equation}
   G_o(\tau):=\left[ G_\uparrow(\tau,\gbf{\kappa}_{11}) \right]_{12}\,.
\end{equation}
Indeed, a change in the PI in our model can be traced back to a sign-change in $G_o(\tau)$ [more precisely, in the corresponding integral of Eq.~(\ref{eq:gomegajl})].
As can be seen from Fig.~\ref{fig:GtauM3_U2}, for $U/t=2$ and $\lambda/t=0.2$, this change occurs  between $t'/t=1.94$ and $t'/t=1.96$, and correspondingly, $\Delta$ jumps from $\Delta=1$ to $\Delta=0$ between these values. This indicates, that for these parameters, the topological-to-trivial band insulator transition occurs for a slightly smaller values of $t'/t=1.95(1)$ than at $U=0$, where the transition takes place at precisely $t'/t=2$. This can be understood to be the consequence of the super-exchange induced by the local Coulomb repulsion which favors the singlet formation on the $t'$-bonds. At the transition point, the single-particle excitation gap $\Delta_{\mathrm{sp}}$ closes, as can be seen from Fig.~\ref{fig:deltasp}, which shows $\Delta_{\mathrm{sp}}$ at the M-point $\gbf{\kappa}_{11}$, obtained from the decay in imaginary time of the diagonal Green's function elements $[G_\uparrow(\tau,\gbf{\kappa}_{11})]_{jj}\propto \exp{(-\tau\Delta_{\mathrm{sp}})}$. This reflects the same gap closing at the transition point as observed for $U=0$ at $t'/t=2$. 

\begin{figure}[t!]
\centering
  \includegraphics[width=\columnwidth]{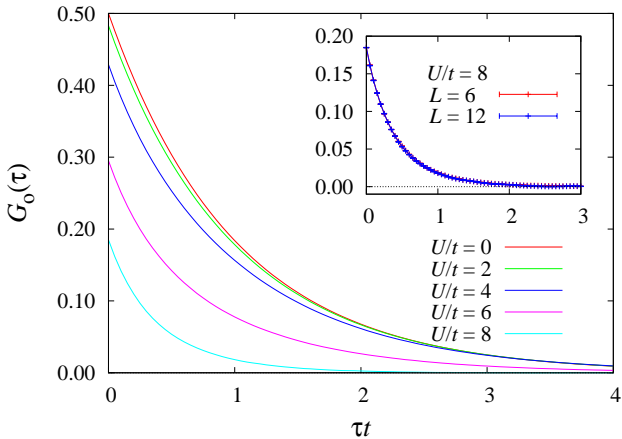}
  \caption{(Color online) Off-diagonal component of the Green's function at the M$_3$-point for $L=6$, $\lambda/t=0.2$, $t'=t$ and different values of $U$ from $U=0$ to $U=8t$ (top to bottom). Across the transition from the QSH insulator to the antiferromagnetic insulator the Green's function remains qualitatively unchanged. Error bars are of the order of the line width and have been omitted for clarity. Inset: The Green's function shows very little finite size dependence due to the large gap induced by the large coupling parameters.
  \label{fig:GtauM3_Uscan}}
\end{figure}

While in the small-$U$ region, the QSH state is stable with respect to interactions and adiabatically connected to the $U=0$ limit, for sufficiently large values of $U$ the system enters a transverse antiferromagnetically ordered Mott-insulating phase, where the time-reversal symmetry of the Hamiltonian $H$ is spontaneously broken in the thermodynamic limit~\cite{Rachel10,Zheng11,Hohenadler11,Hohenadler12a,Assaad13}. This transition is however not related to a closing of the single particle gap, as has been demonstrated by unbiased QMC simulations. The single particle gap only exhibits a local minimum at the transition point, but does not close~\cite{Hohenadler11,Hohenadler12a}. This result from  numerically exact simulations is in contrast to previous VCA calculations, which concluded that the single particle gap closes at the transition to the antiferromagnetic phase~\cite{Yu11}. In fact, the Green's function exhibits no qualitative change across the transition. This can be seen also from the QMC data in Fig.~\ref{fig:GtauM3_Uscan}, where $G_o(\tau)$ is shown for different values of $U$ at $\lambda/t=0.2$ and for $t'=t$.

From previous QMC simulations~\cite{Hohenadler11}, we know that long-ranged antiferromagnetic order sets in for these parameters near $U/t\approx 5$ and flux induced edge states are absent\cite{Assaad13}. However, $G_o(\tau)$  exhibits no significant changes in this interaction region. In particular, and in contrast to the $t'$-scan considered above, $G_o(\tau)$ does not exhibit a change in its sign. That this is not a finite size effect, can be seen in the inset of  Fig.~\ref{fig:GtauM3_Uscan}, where we compare QMC data at $U/t=8$ for two different system sizes, $L=6$ and $L=12$, which are seen to indeed be finite-size converged. We verified that also up to $U/t=40$, no sign change occurs in $G_o(\tau)$. This implies that the PI $\Delta$ stays constant when tuning across the antiferromagnetic transition. We verified explicitly, that even at $\lambda=0$ the PI takes on a non-trivial value in the antiferromagnetic Mott insulating region. 

How does this relate to the quantum phase transition that takes place  when the system enters the antiferromagnetic region, which is thus not adiabatically connected to the $U=0$ state? Only in the thermodynamic limit antiferromagnetic order persists, which spontaneously breaks time-reversal and the inversion (sublattice) symmetry of the Hamiltonian.  Yet this is not monitored by the single particle Green's function, on which the calculation of the PI is based. Spontaneous symmetry breaking in the ordered region implies a degenerate ground state subspace in the thermodynamic limit. In each specific ground state from this manifold, the sublattices A and B are not equivalent anymore, and this condition for a well defined PI is broken. Remarkably, even in the antiferromagnetic region, the degeneracy of the ground state manifold implies the existence of low-energy gapless excitations, namely the Goldstone modes. However, these soft spin excitations are of particle-hole type, and thus not attainable in the single-particle sector.

Interestingly, once the antiferromagnetic order is destroyed in the system by a sufficiently strong dimerization $t'$, the PI does change to a trivial value. To explore this behavior, let us start from the large-$t'$ region, $t'>2t$. Starting at $U=0$ from the  trivial band insulator region for $t'>2t$ and switching on local interactions $U>0$, the system remains insulating, and also does not develop long-range order. This can be most directly seen in the large-$U$ limit. Here, the effective model for the low-energy physics is a Heisenberg model with an exchange dimerization along the $t'$-bonds: The exchange interaction $J'=4t'^2/U$ in second order perturbation theory is more than a factor of 4 larger than the exchange interaction $J=4t^2/U$ along the other nearest-neighbor bonds, and also dominates over the weak (for $\lambda/t=0.2$) next-nearest-neighbor anisotropic exchange coupling $J_2=4\lambda^2/U$ related to the spin-orbit terms~\cite{Rachel10}. The strong $J'$-dimerization drives the spin system into a non-magnetically ordered, dimerized phase. Indeed, for $J_2=0$, the critical ratio beyond which the antiferromagnetic order vanishes in the Heisenberg model on the dimerized honeycomb lattice equals $J'/J=1.735(1)$~\cite{Jiang09}, which relates here to a ratio of $t'/t\approx 1.32$ in the Hubbard model in the large-$U$ limit. For $t'>2t$, the system thus resides inside a non-magnetic phase, adiabatically connected to the trivial band insulator at $U=0$. Correspondingly, the PI of the system does not change upon increasing $U$ at fixed $t'>2t$. 

\begin{figure}[t!]
\centering
  \includegraphics[width=\columnwidth]{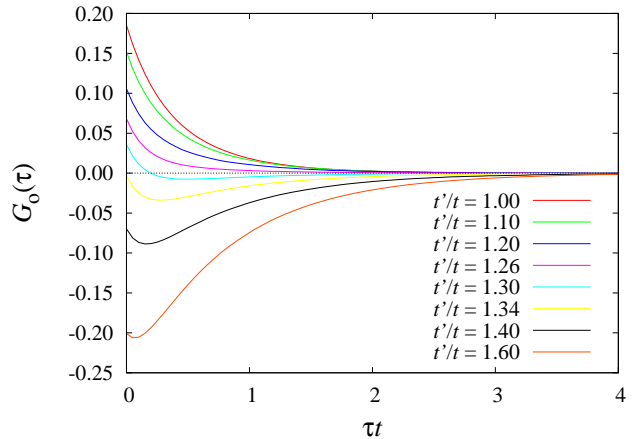}
  \caption{(Color online) Off-diagonal component of the Green's function at the M$_3$-point for $L=6$, $U/t=8$ and $\lambda/t=0.2$ at different values of $t'$ between $t'=t$ and $t'=1.6t$ (top to bottom). Error bars are of the order of the line width and have been omitted for clarity. 
  \label{fig:GtauM3_U8}}
\end{figure}

On the other hand, decreasing the ratio $t'/t$ at sufficiently large $U$, a transition from the large-$t'$ non-ordered phase to the antiferromagnetic phase occurs, and we observe a corresponding transition  in the PI: In Fig.~\ref{fig:GtauM3_U8}, we consider in particular the case of $U/t=8$ and $\lambda/t=0.2$. Upon varying $t'/t$, we find a change in the sign of $G_o(\tau)$, and in more detail, the PI changes from $\Delta=1$ to $\Delta=0$ beyond $t'/t=1.28(2)$. Remarkable is the fact, that this transition is again related to the emergence of low-energy gapless excitations, namely the Goldstone modes, which appear in the antiferromagnetic phase, but not in the dimerized phase, where instead a finite spin-gap separates the singlet ground state from the lowest triplet excited state, that relates in the strong-$J'$ limit to a triplet excitation on one of the strong $J'$ bonds. The single particle gap $\Delta_\text{sp}$ however stays finite in both phases, as well as across the transition.

%------------------------------------------------------------------------------------------------------------
\section{Conclusions}
%------------------------------------------------------------------------------------------------------------
%

We explored Green's function based methods to obtain topological invariants in a two-dimensional strongly interacting fermion system that exhibits  trivial band insulating, Mott insulating and topological insulator regimes. Given an adiabatic connection for a phase of the interacting system to the non-interacting limit, we found that the calculated parity invariant indeed does not change, and thus allows extracting the direct transition between the topological insulator and the trivial insulator region for finite interactions. However, since the parity invariant relates to the single particle Green's function, and hence captures single particle properties only, it does not allow monitoring, e.g., the transition to the antiferromagnetic regime from the topological insulator side. A change of the parity invariant would require corresponding changes in the single particle Green's function, which are not being observed in this case. Use of the parity invariant based on approximate methods to calculate the Green's function  may however lead to deviating conclusions. For example, the change in the parity invariant within the variational cluster approximation to the Kane-Mele-Hubbard model~\cite{Budich12a}, is accompanied by a closing of the gap in the single particle Green's function~\cite{Yu11}, which does not reflect the actual behavior of this model~\cite{Hohenadler11,Hohenadler12a}. Although its usage is thus restricted, we have shown that the parity invariant nevertheless constitutes a readily accessible measure within quantum Monte Carlo simulations for a large variety of (quantum) phase transitions from topological to trivial insulators.

Recently we became aware of a QMC investigation\cite{Hung13} which examined the parity invariant in a related model, focusing on the interaction region below the magnetic ordering transition.

\begin{acknowledgments}
We thank M. Hohenadler, S.~R.~Manmana and M.~J.~Schmidt for discussions and JARA-HPC and JSC J\"ulich for the allocation of CPU time. A. E., V. G. and S. W. acknowledge the Kavli Institute for Theoretical Physics at UCSB for hospitality. This research was supported in part by the National Science Foundation under Grant No. NSF PHY11-25915.
\end{acknowledgments}


\begin{thebibliography}{99}

\bibitem{Hasan10}
M.~Z. Hasan and C.~L. Kane, Rev. Mod. Phys. {\bf 82}, 3045 (2010). 
 
\bibitem{Qi11}
X.-L. Qi, S.-C. Zhang, Rev. Mod. Phys. {\bf 83}, 1057 (2011). 

\bibitem{Bernevig06}
B.~A. Bernevig and S.-C. Zhang, Phys. Rev. Lett. {\bf 96}, 106802 (2006).

\bibitem{Koenig07}
M. K\"onig, S. Wiedmann, C. Br\"une, A. Roth, H. Buhmann, L.~W. Molenkamp, X.-L. Qi, and S.-C. Zhang, Science {\bf 318}, 766 (2007).

\bibitem{Kane05b}
C.~L. Kane and E.~J. Mele, Phys. Rev. Lett {\bf 95}, 226801 (2005).

\bibitem{Kane05a} 
C.~L. Kane and E.~J. Mele, Phys. Rev. Lett. {\bf 95}, 146802 (2005).

\bibitem{Thouless82}
D.~J. Thouless, M. Kohmoto, M.~P. Nightingale, and M. den Nijs, Phys. Rev. Lett. {\bf 49}, 405 (1982).

\bibitem{Fu07}
L. Fu and C.~L. Kane,Phys. Rev. B {\bf 76}, 045302 (2007).

\bibitem{Hughes11}
T.~L. Hughes, E. Prodan, and B.~A. Bernevig, Phys. Rev. B \textbf{83}, 245132 (2011).

\bibitem{Turner12}
A.~M. Turner, Y. Zhang, R.~S.~K. Mong, and A. Vishwanath, Phys. Rev. B \textbf{85}, 165120 (2012).

\bibitem{Hohenadler12b} 
M. Hohenadler and F.~F. Assaad, J. Phys.: Condens. Matter {\bf 25}, 143201 (2013).

\bibitem{Volovik03}
G.~E. Volovik, {\it The Universe in a Helium Droplet}, Oxford University Press, Oxford (2003).

\bibitem{Wang10a}
Z. Wang, X.-L. Qi, and S.-C. Zhang, Phys. Rev. Lett. {\bf 105}, 256803 (2010).

\bibitem{Gurarie11}
V. Gurarie, Phys. Rev. B {\bf 83}, 085426 (2011).

\bibitem{Wang12a}
Z. Wang, X.-L. Qi, and S.-C. Zhang, Phys. Rev. B {\bf 85}, 165126 (2012).

\bibitem{Wang12b}
Z. Wang and S.-C. Zhang, Phys. Rev. X {\bf 2}, 031008 (2012).

\bibitem{Wang12c}
Z. Wang and B. Yan, J. Phys. Condens. Matter {\bf 25}, 155601 (2013). 

\bibitem{Manmanna12}
S.~R. Manmana, A.~M. Essin, R.~M. Noack, and V. Gurarie Phys. Rev. B {\bf 86}, 205119 (2012).

\bibitem{Rachel10}
S. Rachel and K. LeHur, Phys. Rev. B \textbf{82}, 075106 (2010).

\bibitem{Zheng11}
D. Zheng, G.-M. Zhang, and Congjun Wu, Phys. Rev. \textbf{B 84}, 205121 (2011).

\bibitem{Hohenadler11}
M. Hohenadler, T.~C. Lang, and F.~F. Assaad, Phys. Rev. Lett. \textbf{106}, 100403 (2011). 

\bibitem{Wu12}
W. Wu, S. Rachel, W.-M. Liu, and K. Le Hur, Phys. Rev. B \textbf{85}, 205102 (2012).

\bibitem{Hohenadler12a}
M. Hohenadler, Z.~Y. Meng, T.~C. Lang, S. Wessel, A. Muramatsu, and F.~F. Assaad, Phys. Rev. B {\bf 85}, 115132 (2012).

\bibitem{Assaad13}
F.~F. Assaad, M. Bercx, and M. Hohenadler, Phys. Rev. X \textbf{3}, 011015 (2013).

\bibitem{Budich12a}
J.~C. Budich, R. Thomale, G. Li, M. Laubach, and S.-C. Zhang, Phys. Rev. B \textbf{86}, 201407 (2012).

\bibitem{Werner13}
J. Werner, F.~F. Assaad, arXiv:1302.1874 (2013).

\bibitem{Go12}
A. Go, W. Witczak-Krempa, G.~S. Jeon, K. Park, and Y.~B. Kim, Phys. Rev. Lett. {\bf 109}, 066401 (2012).

\bibitem{AsEv08}
F.~F. Assaad and H.~G. Evertz, Lect. Notes Phys. {\bf  739}, 277 (2008).

\bibitem{Meng10}
Z.~Y. Meng, T.~C. Lang, S. Wessel, F.~F. Assaad, and A. Muramatsu, Nature {\bf 464}, 847 (2010).

\bibitem{Sorella12}
S. Sorella, Y. Otsuka, and S. Yunoki, Sci. Rep. {\bf 2}, 992 (2012).

\bibitem{Jiang09}
F.~J. Jiang and U. Gerber, J. Stat. Mech. P09016 (2009). 

\bibitem{Yu11}
S.-L. Yu, X.~C. Xie, and J.-X. Li, Phys. Rev. Lett. {\bf 107}, 010401 (2011).

\bibitem{Hung13}
H.-H. Hung, L. Wang, Z.-C. Gu, and G.~A. Fiete, Phys. Rev. B {\bf 87}, 121113(R) (2013).


\end{thebibliography}
\end{document}